\documentclass[preprint,12pt]{elsarticle}
\usepackage{amssymb}
\usepackage{amsmath}
\usepackage{url}
\usepackage[margin=2.5cm]{geometry}

\makeatletter
\def\ps@pprintTitle{%
 \let\@oddhead\@empty
 \let\@evenhead\@empty
 \def\@oddfoot{\centerline{\thepage}}%
 \let\@evenfoot\@oddfoot}
\makeatother

\begin{document}

\begin{frontmatter}

\title{D-band strain underestimates fibril strain for twisted collagen fibrils at low strains}
\author[inst1,inst2]{Matthew P. Leighton}

\affiliation[inst1]{organization={Department of Physics and Atmospheric Science},
            addressline={Dalhousie University}, 
            city={Halifax},
            postcode={B3H 4R2}, 
            state={Nova Scotia},
            country={Canada}}

\author[inst1]{Andrew D. Rutenberg}
\author[inst1]{Laurent Kreplak}

\affiliation[inst2]{organization={Department of Physics},
            addressline={Simon Fraser University}, 
            city={Burnaby},
            postcode={V5A 1S6}, 
            state={British Columbia},
            country={Canada}}

\begin{abstract}
Collagen fibrils are the main structural component of load-bearing tissues such as tendons, ligaments, skin, the cornea of the eye, and the heart. The D-band of collagen fibrils is an axial periodic density modulation that can be easily characterized by tissue-level X-ray scattering. During mechanical testing, D-band strain is often used as a proxy for fibril strain. However, this approach ignores the coupling between strain and molecular tilt. We examine the validity of this approximation using an elastomeric collagen fibril model that includes both the D-band and a molecular tilt field. In the low  strain regime, we show that the D-band strain substantially underestimates fibril strain for strongly twisted collagen fibrils -- such as fibrils from skin or corneal tissue.
\end{abstract}

\end{frontmatter}


\section{Introduction}
\label{sec:intro}
Collagen-rich tissues such as skin, tendon, ligament, the cornea, and the heart all have complex hierarchical structures that support their load bearing function. The common building block is the collagen fibril, a chiral bundle of collagen molecules whose relative axial stagger gives rise to the D-band --  a characteristic axial density modulation \cite{Orgel:2006}. The D-band can be easily observed by electron microscopy \cite{Hodge:1960}, atomic force microscopy \cite{Fang:2013}, or X-ray scattering \cite{Sasaki:1996}. The D-band repeat is divided in two regions, the overlap region where all the molecules are present in the collagen fibril cross-section and the gap region where 80\% of the molecules are present \cite{Orgel:2006}. In vivo, the enzyme lysyl-oxidase (LOX) cross-links adjacent molecules in the gap region in order to increase tensile stability \cite{Orgel:2001,Makris:2014}. 

Because of their mechanical role, deformation modes of collagen molecules within tissues have inspired experimental and computational research. Two important deformation modes have been identified: stretching mediated by cross-links and intermolecular sliding \cite{Depalle:2015}. (The straightening of `microkinks' along collagen molecules at low strain \cite{Misof:1997} could be seen as a specific stretching mechanism.) Both modes can be observed in stretched tendons over the full range of observable D-band strains using X-ray scattering techniques \cite{Gautieri:2017}. These modes appear to be coupled together at low molecular and D-band strain, though sliding appears to be the dominant mode above $2\%$ fibril strain in tendons \cite{Sasaki:1996, Gautieri:2017}. 

When studying the impact of mechanical deformations on the hierarchical structure of collagen-rich tissues, it is common practice to use changes in the D-band period as a proxy measure of the fibril-level strain \cite{Sasaki:1996, Misof:1997, Aziz:2018, Gautieri:2017, Gachon:2020}. This is based on the untested assumption that changes in the molecular density fluctuation follow fibril elongation in an affine manner. Underlying this assumption are two key conditions: (i) the density of LOX induced cross-links is sufficiently large  and homogeneous to yield a uniform strain field along the collagen fibril, and (ii) changes in collagen molecular orientation due to the axial strain are negligible.

Hydrothermal isometric tension (HIT) relies on the thermal denaturation of collagen-rich tissues at constant tensile strain in order to reveal the connectivity of the network of cross-links within and between collagen fibrils \cite{Lous:1983}. This method confirms that ex vivo collagen tissues are well cross-linked \cite{Lous:1983, Allain:1980, Kampmeier:2000, Herod:2016} -- satisfying the first condition.

For the second condition to be met, the collagen molecular tilt with respect to the fibril axis must be small. This is probably true for tendon fibrils where the molecular tilt at the fibril’s surface is no larger than $5^\circ$ and the D-band spacing is $66-67\mathrm{nm}$ \cite{Hulmes:1981, Quan:2015}. This is unlikely to be true for collagen fibrils in skin and cornea where the molecular tilt at the fibril’s surface can reach $15-20^\circ$ and the D-band spacing is typically $64-65\mathrm{nm}$ \cite{Raspanti:2018, Brodsky:1980}. The inverse relationship between surface molecular tilt and D-band spacing across tissue types can be explained geometrically by considering the projection of a collagen molecule along the fibril axis \cite{Cameron:2020, Bozec:2007}. However, for strained fibrils we must also consider the coupling between tilt and stretch that is often observed in helical assemblies such as actin filaments \cite{Tsuda:1996} or in single helical molecules such as DNA \cite{Sheinin:2009}.

In this work, we use liquid-crystalline elastomer theory to explore the effect of applied strain at the fibril level on both the D-band strain and the molecular tilt. Our collagen fibril model includes a sinusoidal axial density fluctuation giving rise to a global D-band spacing and a double-twist configuration for the molecular tilt \cite{Cameron:2020}. Molecular stretching and intermolecular stretching are implicitly included in the phase-field crystal theory used to model the coupling of D-band deformations with molecular tilt. We limit ourselves to low D-band strains where the D-band energetics are approximately quadratic in strain (see below). Torsion-stretch coupling is observed experimentally at low D-band strain when stretching strips of corneal tissue \cite{Bell:2018} and is well-captured by our model. Stretching the elastomeric fibril shows that molecular tilt decreases rapidly with applied fibril strain while the D-band spacing only increases moderately.  

\section{Continuum theory for a cross-linked collagen fibril}
\label{sec:theory}
We have recently proposed two continuum models for the formation and structure of unstrained collagen fibrils which considered fibril growth either in equilibrium before cross-linking occurs \cite{Cameron:2020} or out of equilibrium where cross-linking occurs during growth \cite{Leighton:2021}. Both models are based on a coarse-grained free energy that accounts for distortions in the molecular orientation field (via a Frank free energy), periodic density modulations (via phase-field crystal theory), and surface effects. The orientation of collagen molecules within a fibril is parametrized by a radius-dependent twist angle $\psi(r)$ at which molecules are tilted with respect to the fibril axis. The molecular director field is thus given by $\hat{\boldsymbol{n}} = -\sin\psi(r) \boldsymbol{\hat{\phi}} + \cos\psi(r) \boldsymbol{\hat{z}}$ in cylindrical coordinates. This `double-twist' is a coarse-grained description of molecular orientation within the collagen fibril \cite{Cameron:2020, Leighton:2021}.

We approximate the D-band as a single-mode sinusoidal density modulation with spacing (wavelength) $d$:
\begin{equation}
\rho(z)  \propto \cos(2 \pi z/d),
\end{equation}
where the $z$ coordinate is aligned along the fibril axis. Increases in wavelength due to extensional strain are expected to occur through the molecular stretching and intermolecular sliding modes mentioned above. Using phase-field crystal theory with this single-mode approximation, the free-energy density averaged over one wavelength is \cite{Cameron:2020, Leighton:2021}
\begin{equation}
\begin{aligned}
f_D & \propto \left(1 - \left( \frac{d_{||}}{d} \right)^2 \cos^2\psi\right)^2,
\end{aligned} \label{dband}
\end{equation}
where $d_\parallel \approx 67\mathrm{nm}$ is the equilibrium (unstrained) D-band period in the absence of molecular twist. Here we only show the energetic contribution from the D-band spacing \cite{Cameron:2020, Leighton:2021}. 

When the molecular twist field $\psi \neq 0$, we cannot minimize the volume-average free-energy using  $d=d_{||}$.  Instead, the D-band spacing that  minimizes the fibril free energy is \cite{Leighton:2021}
\begin{equation}\label{DBandPeriod}
\frac{d}{d_{||}} = \left( \frac{\left\langle \cos^4\psi\right\rangle}{\left\langle \cos^2\psi\right\rangle}\right)^{1/2},
\end{equation}
where angled brackets denote the volume average -- with $\langle \cdot\rangle = 2 \int_0^R \cdot r dr /R^2$, where $R$ is the fibril radius.

Both the D-band period $d_0$ of an unstrained fibril (with $\lambda=1$ and twist angle function $\psi_0(r)$) and the D-band period $d$ of a strained fibril (with strained twist-angle function $\psi(r)$) should satisfy Eqn.~\ref{DBandPeriod}. Accordingly, we can obtain the D-band strain, $\epsilon_D$:
\begin{equation} \label{dbandstraineq}
\epsilon_D \equiv \frac{d-d_0}{d_0}  =\left( \frac{\left\langle \cos^4\psi\right\rangle\left\langle \cos^2\psi_0\right\rangle}{\left\langle \cos^2\psi\right\rangle\left\langle \cos^4\psi_0\right\rangle}\right)^{1/2}-1, 
\end{equation}
where we express $\epsilon_D$ solely in terms of the strained and unstrained twist-angle functions. Note that at small $\epsilon_D$ the D-band energetics ($f_D$ in Eqn.~\ref{dband}) are quadratic in $\epsilon_D$. We expect that $f_D$ requires additional higher order terms in $\epsilon_D$ at larger strains. Accordingly our results are restricted to sufficiently small $\epsilon_D$. 

When the fibril is strained axially, we can apply elastomeric theory \cite{Warner:1996} to the cross-linked fibril to determine the strained twist function $\psi(r)$ from  the fibril stretch ratio $\lambda$ and the initial twist function $\psi_0(r)$. Assuming that the elastomeric free energy dominates the Frank and D-band free energies \cite{Leighton:2021}, we can minimize the elastomeric free energy to determine the equilibrium strained configuration. This minimization can be performed analytically \cite{Leighton:2021b}, and yields
\begin{equation}\label{psieq}
    \psi(r) = \frac{1}{2}\cot^{-1}\left( \frac{ (\zeta+1)(\lambda^3-1) + (\zeta-1)(\lambda^3+1)\cos(2\psi_0)}{2\lambda^{3/2}(\zeta-1)\sin(2\psi_0)} \right)
\end{equation}
The parameter $\zeta$ quantifies the anisotropy of the intermolecular cross-links; it is defined as the ratio between the mean lengths of cross-links in the directions parallel and perpendicular to the molecular director field. 
Any value of $\zeta\geq0$ could in theory be realized in an elastomer system, however $\zeta>1$ is generally assumed in modelling approaches \cite{Warner:1996,Warner:2000} and observed experimentally for nematic liquid crystals \cite{DAllest:1988,Kundler:1998}. For extensional fibril strains (corresponding to $\lambda>1$) we predict that the molecular twist $\psi$ decreases (increases) when $\zeta>1$ ($\zeta<1$). Since molecular twist of collagen molecules has been observed experimentally to decrease when fibrils are axially extended \cite{Bell:2018}, we restrict ourselves to $\zeta>1$. 

To summarize, for a given unstrained twist angle function $\psi_0(r)$, we can compute as a function of the fibril strain $\epsilon_F = \lambda-1$ the post-strain twist angle function $\psi(r)$ (using Eqn.~\ref{psieq}) and the D-band strain $\epsilon_D$ (using Eqn.~\ref{dbandstraineq}). These calculations depend only on a single parameter, the cross-link anisotropy $\zeta$. We can compute solutions to Eqns.~\ref{dbandstraineq} and \ref{psieq} numerically for arbitrary $\psi_0$, $\zeta$, and $\epsilon_F$ using code we have made publically available on GitHub \cite{github}. 

The modelling framework presented above is valid at small D-band strain and so also small fibril strain. Based on geometrical considerations, deformation via molecular untwisting can only accommodate fibril strains up to  
\begin{equation} \label{e_untwist_max}
\epsilon_\mathrm{untwist}^\mathrm{max} \approx \sec\left[\psi_0(R)\right]-1,
\end{equation}
where $\psi_0(R)$ is the surface twist of an unstrained fibril. For corneal fibrils which have a surface twist of about $18^\circ$, this would limit the applicability of our model to fibril strains less than or equal to $5-6\%$. 

\subsection{Small angle limit}
Under extension (with $\lambda>1$) and with small initial twist angles the strained twist angle function Eqn.~\ref{psieq} is approximately given by \cite{Leighton:2021b}
\begin{equation}\label{smallangle}
\psi(r) \simeq \psi_0(r) (\zeta-1)\left(\zeta\lambda^{3/2} - \lambda^{-3/2}\right)^{-1}.
\end{equation}
We can use this approximation to obtain a small-angle approximation for the D-band strain $\epsilon_D$ as a function of the initial and strained volume-averaged twist angle functions:
\begin{equation}\label{DbandStrain}
\epsilon_D \simeq \frac{1}{2}\left[ 1 - \left( \frac{\langle \psi\rangle}{\langle\psi_0\rangle}\right)^2\right]\left\langle \psi_0^2\right\rangle.
\end{equation}
Alternatively, we can write $\epsilon_D$
in terms of the fibril stretch ratio $\lambda= 1 + \epsilon_F$ as
\begin{equation}\label{DbandvsFibril}
\epsilon_D \simeq \frac{1}{2}\left[ 1 - \left( \frac{\zeta - 1}{\zeta\lambda^{3/2} - \lambda^{-3/2}}\right)^2\right]\left\langle \psi_0^2\right\rangle.
\end{equation}

From Eq. \ref{DbandStrain} we see that $\epsilon_D$ increases monotonically as $\langle\psi\rangle$ decreases due to molecules untwisting. This function is concave ($\partial^2 \epsilon_D/\partial \langle\psi\rangle^2\leq0$), so that the D-band strain grows faster when the molecular twist is higher. For large strains the molecular tilt will disappear ($\psi\to 0$), and thus the D-band strain asymptotically approaches the value $\langle\psi_0^2\rangle/2$ in the small angle limit, and $\left[\langle\cos^2\psi_0\rangle/\langle\cos^4\psi_0\rangle\right]^{1/2} - 1$ more generally. 

Eq.~\ref{DbandvsFibril} has simple asymptotic behavior as a function of $\zeta$. When  $\zeta$ approaches $1$ from above, the D-band strain is 
\begin{equation}
\epsilon_D = \begin{cases} 
0, & \epsilon_F=0\\
\frac{1}{2}\langle\psi_0^2\rangle, & \epsilon_F>0.
\end{cases}
\end{equation}
When $\zeta\to\infty$ we have $\epsilon_D = \frac{1}{2}\left(1 - \lambda^{-3}\right)\langle\psi_0^2\rangle$. In the small angle limit the D-band strain is always a monotonically increasing ($\partial \epsilon_D/\partial\epsilon_F \geq0$) and concave ($\partial^2 \epsilon_D/\partial\epsilon_F^2 \leq0$) function of fibril strain, as long as $\zeta>1$ and $\epsilon_F>0$.  Monotonicity and concavity also hold (not shown) for the more general D-band strain function given by Eq.~\ref{dbandstraineq}.

\subsection{D-band strain vs fibril strain}\label{cornealcompare}
Our small-angle approximations in Eqns.~\ref{DbandStrain} and \ref{DbandvsFibril} are useful for comparing with experimental data. Even for collagen fibrils with relatively high surface twist, such as the $\psi_0(R)\approx 0.3$ observed in corneal fibrils, the small angle approximation applies since we expect that the twist-angle function monotonically decreases from the surface \cite{Cameron:2020, Leighton:2021}. In general, the full twist angle function $\psi_0(r)$ is unknown, and only the surface twist $\psi_0(R)$ and the volume-averaged twist $\langle \psi_0\rangle$ can be measured. Accordingly, we treat $\langle\psi_0^2\rangle$ as an adjustable parameter.

\begin{figure}[h] 
\centering
  \includegraphics[width=\textwidth]{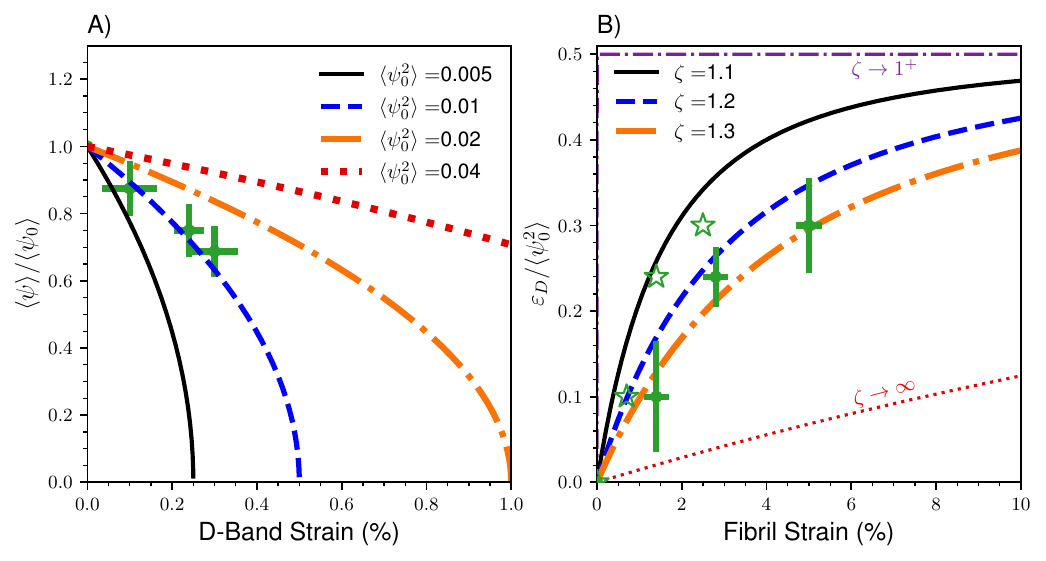}
  \caption{A) The scaled volume-averaged twist-angle $\langle\psi\rangle/\langle\psi_0\rangle$ vs. D-band strain $\epsilon_D$, for the indicated values of $\langle\psi_0^2\rangle$ (lines). B) The D-band strain, scaled by $\langle\psi_0^2\rangle$, vs. fibril strain $\epsilon_F$ for indicated values of $\zeta$ (lines). The asymptotic limits of $\zeta \rightarrow 1^+$ and $\zeta \rightarrow \infty$ are indicated by thin dot-dashed and dotted lines, respectively. In both panels we compare our predictions with data from \cite{Bell:2018} (green crosses). In B) we have scaled the measured D-band strain by our estimate $\langle\psi_0^2\rangle\approx0.01\mathrm{rad}^2$, and assumed that the fibril strain is either equal to the reported tissue strain (green crosses) or half of it (unfilled green stars).}
  \label{fig:dbandstrain}
\end{figure}

Fig.~\ref{fig:dbandstrain}A shows how the volume-averaged twist angle decreases with D-band strain in the small-angle limit (Eqn.~\ref{DbandStrain}), for different values of $\langle\psi_0^2\rangle$. We also show experimental data from \cite{Bell:2018} (green crosses), which reported measurements of both D-band strain and corresponding changes in volume-averaged molecular tilt for strained corneal tissue. We find that our model  fits the data with $\langle\psi_0^2\rangle\approx 0.01\mathrm{rad}^2$. Note however that the measured D-band strain may be an underestimate due to averaging over a broad distribution of fibril angles \cite{Bell:2018}.

Fig.~\ref{fig:dbandstrain}B then shows D-band strain vs fibril strain curves (as given by Eqn.~\ref{DbandvsFibril}) for different values of $\zeta$. Different values of $\zeta$ can lead to a variety of curves. We can compare with experimental data from \cite{Bell:2018} here as well, which we show with green crosses (measurements of D-band strain are  scaled by our estimated value of $\langle\psi_0^2\rangle$). Under the assumption that the fibril strain is equal to the reported tissue strain, we find good agreement with our model predictions for $\zeta\approx1.3$. Fibril strain could be smaller than tissue (applied) strain if fibrils are not perfectly aligned with the applied strain -- or if there is  slippage between fibrils \cite{Gupta:2010, Puxkandl:2002}. For example, we have also shown the assumption that fibril strain is half of the reported strain (unfilled green stars). Accordingly, we expect $\zeta\in[1,1.3]$ for corneal fibrils. An additional datapoint at $8\%$ tissue strain -- but no measured D-band strain-- is not shown, as it is beyond our approximate valid range of tissue strain (see above). Consistent with this, the additional datapoint would be $\epsilon_D/\langle \psi_0^2 \rangle \approx 0.59$ -- which is above the maximal $\zeta \rightarrow 1$ limit of $0.5$ given by our theory. 

We can use Eq.~\ref{DbandvsFibril} to compare the D-band and fibril strains. At leading order (for small extensional strains) we obtain 
\begin{equation}
    \epsilon_D \approx \frac{3}{2} \frac{\zeta+1}{\zeta-1} \langle \psi_0^2 \rangle \, \epsilon_F.
\end{equation}
The D-band strain will underestimate the fibril strain whenever $\langle\psi_0^2\rangle <(2/3) (\zeta-1)/(\zeta+1)$ -- i.e. at small tilt angles. More generally we do not expect to observe $\epsilon_D$ equal to $\epsilon_F$. 

\subsection{D-band energetics}
We have assumed that the D-band phase-field term ($f_D$ in Eqn.~2) is energetically subdominant to elastomeric energies. This assumption is needed to be able to use the elastomeric theory to determine $\psi(r)$ (via Eqn.~\ref{psieq}), with the D-band strain then determined from $\psi$ using Eqn.~\ref{dbandstraineq}. We can estimate the amplitude of $f_D$ from observed variations of the D-band spacing if we assume that they correspond to equilibrium fluctuations. From Eqn.~\ref{dband} and including constant factors we have \cite{Cameron:2020}
\begin{equation}
    \frac{\sigma^2_d}{d^2} = \frac{k_B T}{\pi R^2 d Y_D},
\end{equation}
where the left-side is the fractional variance of D-band spacing, while the right-side includes $k_B$ Boltzmann's constant, the temperature $T$, fibril radius $R$, and a proportionality factor determining a Young's modulus $Y_D$. 

Using $\sigma^2_d/d^2 \simeq 10^{-4}$ and $R \simeq 50\mathrm{nm}$ for collagen fibrils \cite{Fang:2012,Fang:2013B}, $k_B T \simeq 4.1 \mathrm{pN nm}$ and $d \simeq 67\mathrm{nm}$, we estimate $Y_D \simeq 80 \mathrm{kPa}$. This is much smaller than the Young's modulus $Y \gtrsim 100\mathrm{MPa}$ measured for cross-linked collagen fibrils \cite{Graham:2004}. Since $Y$ determines the scale of the mechanical energy at small strains, we confirm that the D-band energetics are subdominant to the elastomeric energetics at small strains. 

\section{Discussion}
\label{sec:discussion}
Using an elastomeric model for cross-linked double-twist fibrils, combined with a subdominant phase-field model for axial (D-band) modulations that are coupled with the double-twist, we have shown how the D-band strain $\epsilon_D$ is typically much less than the fibril strain $\epsilon_F$ under small axial extensions. We have limited our results to the small-angle regime appropriate for collagen fibrils, but our results are qualitatively similar for larger twist angles. In validation, we have shown how experimental data on collagen fibrils from corneal tissue are well fit by our results. 

Experimentally, measuring changes in D-band spacing requires small angle X-ray scattering \cite{Sasaki:1996, Gautieri:2017} (SAXS) while measuring molecular tilt can be achieved by analyzing the angular dependency of the radial molecular spacing using wide angle X-ray scattering (WAXS) \cite{Bell:2018}. With our model, it is now possible to fit the relationship between average twist ($\langle \psi \rangle$) and D-band strain ($\epsilon_D$) to estimate the true fibril strain ($\epsilon_F$). In the process, we also estimate two new quantities for corneal collagen fibrils: the cross-linking anisotropy parameter $\zeta \simeq 1.3$ and the volume-average square of the tilt-angle $\langle \psi_0^2 \rangle \simeq 0.01$. 

As can be seen in Fig.~\ref{fig:dbandstrain}B, mechanical properties of fibrils are highly sensitive to the cross-link anisotropy parameter $\zeta$; consistent with previous theoretical treatment of the mechanics of cross-linked fibrils \cite{Leighton:2021b}. While we do not know the anisotropy parameters for advanced glycation end-product (AGE) vs. enzymatic (LOX) crosslinks, we do expect them to differ since $\zeta$ will depend on the molecular details. Other structural changes such as mineralization may also impact the cross-link anisotropy.  Our results, with sensitive $\zeta$ dependence, may prove useful to assess cross-linking or other structural changes in fibrils. Conversely, manipulating $\zeta$ experimentally may prove to be a practical avenue to tune aspects of the mechanical response of collagen fibrils.

A similar tilt-stretch coupling should also occur in skin where the D-band spacing is $65\mathrm{nm}$ \cite{Brodsky:1980} and the surface molecular tilt reaches $17^\circ$ \cite{Ottani:2001, Mechanic:1987}. Notch testing has shown that skin resists tearing by stretching the fibrils perpendicular to the propagation direction of the tear \cite{Yang:2015}. For these fibrils the D-band spacing increases up to $67\mathrm{nm}$ before failure starts to occur at $3\%$ D-band strain \cite{Yang:2015}. This limiting D-band spacing equals the unstrained value observed in tendon \cite{Quan:2015}.  $\langle \psi \rangle$ has not yet been measured in this tissue. Nevertheless, our results indicate that the $3\%$ D-band strain at failure may correspond to significantly larger fibril strains. 

When the surface molecular twist is smaller than $5^\circ$, as in tendon fibrils \cite{Hulmes:1981}, strain-straightening of molecules is not a significant deformation mechanism, and molecular stretching combined with intermolecular sliding dominates the response \cite{Sasaki:1996, Gautieri:2017}. Indeed, we expect the level of strain that can be accommodated by the molecular untwisting mechanism described here to be no more than $\epsilon_\mathrm{untwist}^\mathrm{max}$ as defined in Eq.~\ref{e_untwist_max}. Other collagenous tissues, including bone \cite{Almer:2005, Zimmermann:2011, Xi:2020} and cartilage \cite{Inamdar:2019}, that exhibit smaller unstrained D-bands indicating significant molecular twist should also exhibit torsion-stretch coupling at small strains up to $\epsilon_\mathrm{untwist}^\mathrm{max}$. While this limit can be as high as $5-6\%$ strain in highly twisted corneal tissue, it is less than half a percent for the small molecular twist of tendon fibrils.  We expect that the torsion-stretch coupling can be mostly ignored in tendon; so that D-band strain should remain a good measure of fibril strain in tendon.

Given the variety of collagen fibril surface tilts observed, it is interesting to consider the possible functional advantages of having a larger molecular tilt. One possibility is that the molecular tilt acts as a reversible deformation mechanism at low fibril strains -- much as the low-stiffness toe-region of tissue elasticity arises from crimp removal \cite{Fratzl:1998}. This could protect structured collagenous tissue in the low-strain regime. An additional example of this protective mechanism could be in the chordae that control the position of valve leaflets in the heart \cite{Ross:2020}. The chordae contain both collagen fibrils and elastin fibres arranged in a multi-layered cylindrical structure \cite{Millington:1998}. Static measurements show that the collagen fibrils have a D-band spacing of $65\mathrm{nm}$ and an average molecular twist angle of $9^\circ$ or $0.15$ radians \cite{Folkhard:1987}. This is half the value observed in the cornea, but significantly larger than in tendon.

\section{Conclusions}
\label{sec:conclusion}
While the molecular twist of a collagen fibril is difficult to observe experimentally, our model shows that twist can have a large impact on the elastic properties of a fibril. We have shown that torsion-stretch coupling leads to D-band strains substantially smaller than small fibril strains. This torsion-stretch coupling could enable fibrils in the cornea, the chordae and the skin to delay the onset of plastic deformation that can occur at small D-band strains \cite{Gautieri:2017}. Significant molecular tilt in these tissues may have evolved to increase their resistance to damage due to cyclic loading.

While the direct applications of our results are to collagen fibrils, they should also apply to other double-twisted filaments with axial modulations such as keratin macrofibrils in hair and wool \cite{Kreplak:2002, Harland:2014}. 

\section*{Conflicts of interest}
There are no conflicts to declare.
\section*{Acknowledgements}
We thank the Natural Sciences and Engineering Research Council of Canada (NSERC) for operating Grants RGPIN-2018-03781 (LK) and RGPIN-2019-05888 (ADR). MPL thanks NSERC for summer fellowship support (USRA-552365-2020), and a CGS Masters fellowship.

\bibliographystyle{elsarticle-num} 
\bibliography{main}
\end{document}